# Magnetic Moment of Leptons

Samina Masood* and Holly Mein
Department of Physical and Applied Sciences,
University of Houston-Clear Lake, Houston TX 77058
*E.mail: masood@uhcl.edu

**Abstract**

We show that the magnetic moment of leptons is significantly modified in thermal background as compared to the corresponding vacuum value. We compare the magnetic moment of all different leptons near nucleosynthesis. It is shown that the significance of thermal corrections depends on the temperature of the universe and the respective lepton mass. In the early universe, particle mass was growing quadratically with temperature which affects the corresponding value of magnetic moment. Intrinsic magnetic moment is inversely proportional to mass whereas thermal corrections to the neutrino dipole moment is proportional to neutrino mass for the Dirac type neutrino in the minimal standard model. Therefore the effect of temperature is not the same for charged leptons and neutrinos.

## 1. Introduction

Magnetic moment is an intrinsic property of charge and depends on mass explicitly. Thus it is affected by the change in physical measurement of mass [1-3, 11-20] and charge [19-23] of particles at high temperatures of the early universe. Temperature dependence of electron mass and charge has been calculated using the renormalization techniques of quantum electrodynamics (QED) at finite temperature [4-8, 13-15]. In this paper we compare the selfmass and magnetic moment of different lepton species at finite temperatures [5-8, 24-28]. The charge of electrons and the electromagnetic coupling quadratically depend on the temperature and leads to the change in electromagnetic properties [24-34] of the medium.

When a particle interacts with a magnetic field, a force is exerted on the particle producing a torque or the magnetic moment. Quantum mechanically, the magnetic moment is expressed in units of Bohr magneton $\mu_B$, whereas intrinsic angular momentum is called particle spin and is proportional to $\hbar$.

Charged leptons have an intrinsic magnetic moment due to the interaction with the surrounding charges, whereas the corresponding massless neutrinos have zero magnetic moment. In other words, nonzero mass of neutrino leads to nonzero value of the magnetic moment of neutrino. We calculate the magnetic moment of neutrinos in a minimal standard model to compare the relative contribution of magnetic moment. Minimal standard model refers to the standard model which requires the conservation of individual lepton number, but the right-handed neutrino is added as a singlet to each generation to accommodate the massive neutrino.

Magnetic moment of a charged particle is an important quantity to study the interaction of particles with their surroundings. We use the physically measurable values of QED parameters which have been evaluated in literature using the renormalization scheme of QED in real-time formalism [10-11]. The masses of neutrinos depend on the choice of models. We use minimal standard model where all the conservations rules of standard model including individual lepton number conservation are still obeyed.

One Bohr magneton $\mu_B$ is equal to the intrinsic magnetic moment of an electron in units of the magnetic field. $m_e$ is the mass of electron.

$$\mu_B = \frac{e\hbar}{2m_e} \quad (1)$$

The magnetic moment of any charged lepton flavor ` $\ell$ ' (with $\ell$ = e, μ, τ) can be related to the Bohr magneton as:

$$\mu_\ell = \frac{m_e}{m_\ell}\mu_B \quad (2)$$

In the next section of this paper we discuss the lepton magnetic moment in vacuum. Section 3 is comprised of the calculations of the magnetic moment up to the first order in alpha and checks their relevance for different temperatures in the early universe. The discussion of the results is included in the last section.

## 2. Magnetic Moment of Leptons in Vacuum

The magnetic moment of an electron is considered intrinsic magnetic moment as it is an intrinsic property of charge. This happens when the spin of a particle interacts with an external magnetic field and depends on the mass [14] of the corresponding lepton. These calculations for charged leptons in a vacuum show that the mass is inversely proportional to the magnetic moment. The larger the mass, the smaller the magnetic moment will be.

The induced magnetic moment is a relatively distinct feature of neutrinos which are very light in mass and are electrically neutral. The intrinsic magnetic moment of neutrino is essentially zero. However, the extremely light neutrinos may show some indirect effect of induced magnetic moment in measurable quantities. Therefore, the calculation of magnetic dipole moment of neutrino in terms of the corresponding leptons is worth-doing. This relationship between charged lepton and neutrino can be seen in the Feynman diagrams, given in Figs. (1) and (2).

We consider the Dirac mass of neutrino where the neutrino and antineutrino satisfy the Dirac equation. Magnetic moment of the Dirac neutrino was first calculated by Lee and Shrock [13]. Their calculation gave the magnetic moment in terms of the mass of neutrino (expressed in electron volt), substituted in units of $\mu_B$:

$$\mu_\ell^D = \frac{3eG_F m_{\nu_\ell}}{8\sqrt{2}\pi^2} \approx 3.2 \times 10^{-19} \left(\frac{m_{\nu_\ell}}{1eV}\right)\mu_B \quad (3)$$

$$G_F = \frac{\sqrt{2}}{8}\frac{g^2}{m_W^2 c^4} = 1.166 \times 10^{-5} GeV^{-2} \quad (4)$$

$G_F$ is the Fermi Coupling Constant in the weak interaction and $m_{\nu_\ell}$ is the mass of the corresponding flavor of neutrino. The magnetic moment is very small compared to the charged lepton of the same flavor.

We generalize Eq. (3) to calculate the magnetic moment of different neutrino species in vacuum. We use the most up to date masses that have been provided experimentally [31, 34]. These may be different from the masses that are used in previous papers on the magnetic dipole moment of the neutrino because more exact values of the neutrino have been found experimentally.

We generalize Shrock's calculation for the magnetic moment of an electron neutrino in a vacuum to all neutrino flavors as:

$$\mu_{\nu_\ell}^0 = \frac{3eG_F m_{\nu_\ell}}{8\sqrt{2}\pi^2} = \frac{3m_e G_F m_{\nu_\ell}}{4\sqrt{2}\pi^2} (\mu_B) \approx 3.2 \times 10^{-19} \left(\frac{m_{\nu_\ell}}{1eV}\right) \mu_B \qquad (5)$$

$m_{\nu_\ell}$ denotes mass of neutrino. Magnetic moment of each neutrino flavor can be evaluated from this equation and is listed in Table 2.

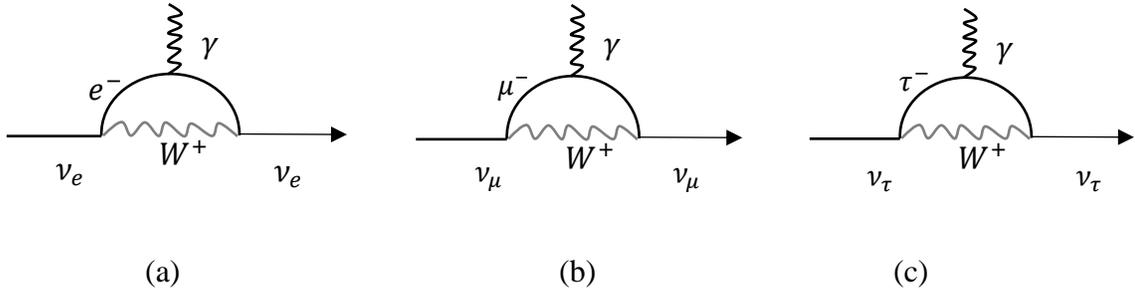

(a) (b) (c)

**Fig.1:** Bubble diagrams for different flavors of neutrinos in the minimal standard model. (a) corresponds to electron-neutrino (b) muon-neutrino and (c) tau-neutrino

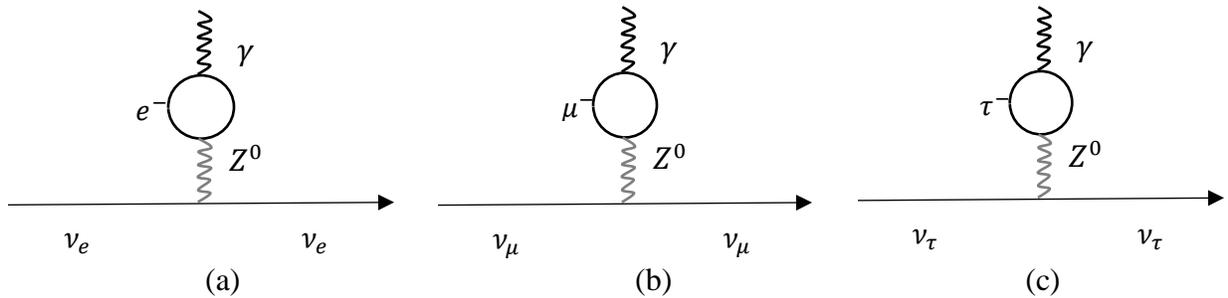

(a) (b) (c)

**Fig.2:** Tadpole diagrams for different flavors of neutrinos in the minimal standard model. (a) corresponds to electron-neutrino (b) muon-neutrino and (c) tau-neutrino

## 3. Magnetic Moment at Finite Temperature

The lepton selfmass affects the intrinsic magnetic moment of the corresponding charged lepton. The renormalization scheme of QED has been used in real-time formalism showing the order-by-order cancellation of singularities (KLN-theorem [10]). The contribution of the thermal background can be evaluated in real-time formalism [1-3, 14-16]. Using the lepton masses given in Section 2, the renormalized mass of the leptons can be written as:

$$m^{phys}_\ell = m_\ell + \delta m_\ell \qquad (6)$$

Eq. (6) is a straightforward generalization of the calculation of electron self-mass [32]. $m_\ell$ represents the bare mass of lepton with the flavor $\ell$.

The first order temperature correction for the electron mass at a finite temperature is equal to:

$$\frac{\delta m_\ell}{m_\ell} \simeq \frac{1}{2m_\ell^2}\{(m^{phys}_\ell)^2 - m_\ell^2\})$$

$$\simeq \frac{\alpha \pi T^2}{3m_\ell}\left[1 - \frac{6}{\pi^2}c(m_\ell\beta)\right] + \frac{2\alpha}{\pi}\frac{T}{m_\ell}a(m_\ell\beta) - \frac{3\alpha}{\pi}b(m_\ell\beta) \qquad (7)$$

For $\alpha = \frac{e^2}{4\pi\hbar c} \simeq \frac{1}{137}$, the coefficients of $a(m_\ell\beta)$, $b(m_\ell\beta)$, and $c(m_\ell\beta)$ contribute differently at different temperatures corresponding to the lepton flavors $\ell$. At low temperatures these coefficients are unimportant and can be dropped off, giving selfmass to be:

$$\frac{\delta m_\ell}{m_\ell} \xrightarrow{T \ll m_\ell} \frac{\alpha \pi T^2}{3m_\ell^2} \qquad (8a)$$

The coefficients a, and b are very small and ignorable at high temperature, and c is summed to $-\frac{\pi^2}{12}$, giving:

$$\frac{\delta m_\ell}{m_\ell} \xrightarrow{T \gg m_\ell} \frac{\alpha \pi T^2}{2m_\ell^2} \qquad (8b)$$

Eq. (7) is relevant for calculations pertaining to primordial nucleosynthesis which occurs near the electron mass temperature. Eqs. (7, 8) are used to estimate thermal contribution to the magnetic moment of all leptons in the standard model. Additionally, above equations help to determine thermal contributions to all leptons expressed in units of Bohr magneton.

The radiative corrections to the lepton selfmass in a thermal background of fermions and bosons significantly contribute to the magnetic moment of leptons. It has been previously shown that the selfmass of particles contribute to the magnetic moment (in units of $\mu_B$) as:

$$\mu_a = \frac{\alpha}{2\pi} - \frac{2}{3}\frac{\delta m_\ell}{m_\ell} \qquad (9)$$

The renormalizability of QED establishes the fact that one-loop thermal corrections dominate over higher order corrections. In this paper, we restrict ourselves to the first order thermal corrections only. Therefore, the one-loop thermal corrections to the magnetic moment of leptons in terms of the corresponding coefficients $a(m_\ell\beta)$, $b(m_\ell\beta)$ and $c(m_\ell\beta)$ [3] are given as:

$$\mu^\beta_\ell = -\frac{2\alpha\pi T^2}{9m_\ell}\left[1 - \frac{6}{\pi^2}c(m_\ell\beta)\right] + \frac{2\alpha}{\pi}\frac{T}{m_\ell}a(m_\ell\beta) - \frac{3\alpha}{\pi}b(m_\ell\beta) \qquad (10)$$

Whereas the net values of lepton magnetic moment with thermal corrections is:

$$\mu_\ell = \frac{\alpha}{2\pi}\left(1 - \frac{2}{3}\frac{\delta m_\ell}{m_\ell}\right)\left(\frac{m_e}{m_\ell}\right)\mu_B \qquad (11)$$

At temperatures sufficiently below the lepton mass, thermal contributions of the corresponding fermions are stifled in the standard model and only photons contribute to the temperature dependent corrections. However, for T<< $m_\ell$, Eq. (11) leads to:

$$\mu_a = \frac{\alpha}{2\pi} - \frac{2}{9}\frac{\alpha\pi T^2}{m_\ell^2} \quad (12a)$$

and for T >> $m_\ell$:

$$\mu_a = \frac{\alpha}{2\pi} - \frac{1}{3}\frac{\alpha\pi T^2}{m_\ell^2} \quad (12b)$$

The next step is changing the mass into units of Kelvin for c=k$_B$=1, where k$_B$ is the Boltzmann constant.

Thermal contributions to the magnetic moment of electron-neutrino before nucleosynthesis [16] at T >> $m_e$ is:

$$\mu_e = 1.17 \times 10^{-3} \left(1 - \frac{2\pi^2 T^2}{3m_e^2}\right)\mu_B \quad (13a)$$

Whereas, for T << $m_e$, the magnetic moment value comes out to be:

$$\mu_e = 1.17 \times 10^{-3} \left(1 - \frac{4\pi^2 T^2}{9m_e^2}\right) \quad (13b)$$

We see this thermal contribution is greater than the corresponding vacuum value of the magnetic moment at high temperatures causing a flip in the magnetic moment.

Muons and tauons have similar behavior as electron in thermal background. In this case, the photons interact through muon and tauon loops instead of electron loop due to the individual flavor conservation. For this purpose, we generalize the existing results of electron to all lepton flavors in the early universe. However, the temperature correspondence with the lepton mass will vary with the lepton flavor and is mentioned in Tables 1 and 2.

Thermal contributions to the magnetic moment of neutrinos are calculated from the bubble diagrams of Fig. (1), and they are generalized to all lepton flavors as:

$$a_{\nu_\ell} = \frac{T^2 G_F m_e m_{\nu_l}}{12M^2}\mu_B \quad (14)$$

Thermal contributions to the magnetic moment of neutrino at the nucleosynthesis temperatures are given below to show that this contribution is larger for heavier neutrinos as compared to the lighter ones, in the standard model.

$$a_{\nu_e} = \frac{T^2 G_F m_e m_{\nu_e}}{12M^2}\mu_B \approx 6.58 \times 10^{-16}\mu_B \quad (15a)$$

$$a_{\nu_\mu} = \frac{T^2 G_F m_e m_{\nu_\mu}}{12M^2}\mu_B \approx 5.55 \times 10^{-11}\mu_B \quad (15b)$$

$$a_{\nu_\tau} = \frac{T^2 G_F m_e m_{\nu_\tau}}{12M^2}\mu_B \approx 4.53 \times 10^{-9}\mu_B \quad (15c)$$

It is also worth-mentioning that we include thermal corrections to mass and the magnetic moment, but keep the charge and the coupling constant independent of temperature. It will not significantly affect the results as the thermal corrections to charge and the QED coupling constant are much smaller than the corresponding corrections to mass and the magnetic moment.

### 4. Discussion

Leptons were the first major particles produced in the universe and a detailed understanding of their behavior is expected to resolve unsolved mysteries of the universe. High temperatures of that early universe must have played a crucial role in the properties of particles. Lepton mass and charge are known to be modified in QED plasma [21-27], at that time, through vacuum polarization. Electromagnetic properties of particles in the early universe are also determined from mass, charge and the magnetic moment of leptons.

Implications of thermal corrections to the magnetic moment are relevant in the early universe especially at very high temperatures. The magnetic moment of charged leptons is inversely proportional to the square of its mass, so the heavier leptons have much smaller magnetic moment as compared to electron (Eq. (13)). Thermal contributions to magnetic moment of neutrinos are proportional to neutrino mass (Eq. (14)) and thermal corrections to heavier neutrinos contribute more significantly than lighter ones. However, the heavier neutrinos can only exist at higher energies.

Thermal contribution to lepton magnetic moment is subtracted from the magnetic moment in vacuum (Eq. (11)) reducing its net value with the increase in temperature. It is demonstrated by the fact that thermal contribution to electron magnetic moment is around 0.7% whereas the contributions to the muon or tauon is much smaller and ignorable, as shown in Table 1. However, at higher temperatures near the tauon mass, the magnetic moment of the higher mass particles will not remain ignorable. We have plotted the magnetic moment of electron, muon and tau leptons as a function of temperature in Figs. (3). Temperature dependence is similar whereas the magnitude of thermal contribution is suppressed with the square of mass. All of these magnetic moments are plotted in units of Bohr magneton.

**Figure 3a: Electron Lepton Magnetic Moment vs $\left(\frac{T}{m}\right)^2$**

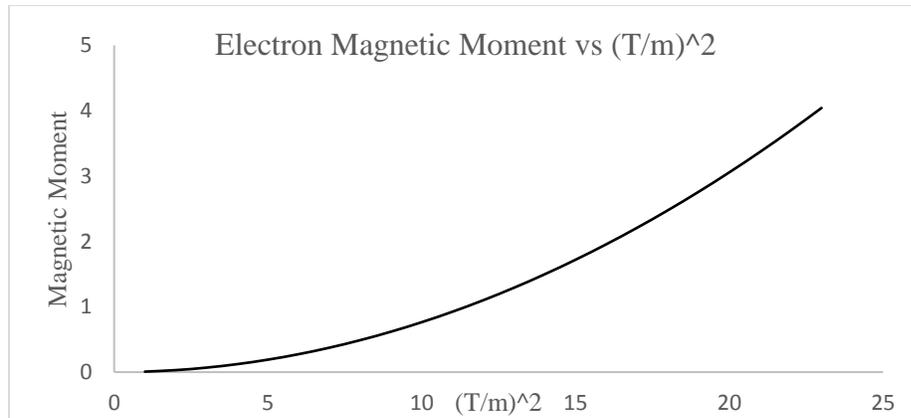

**Figure 3b: Muon Lepton Magnetic Moment vs $\left(\frac{T}{m}\right)^2$**

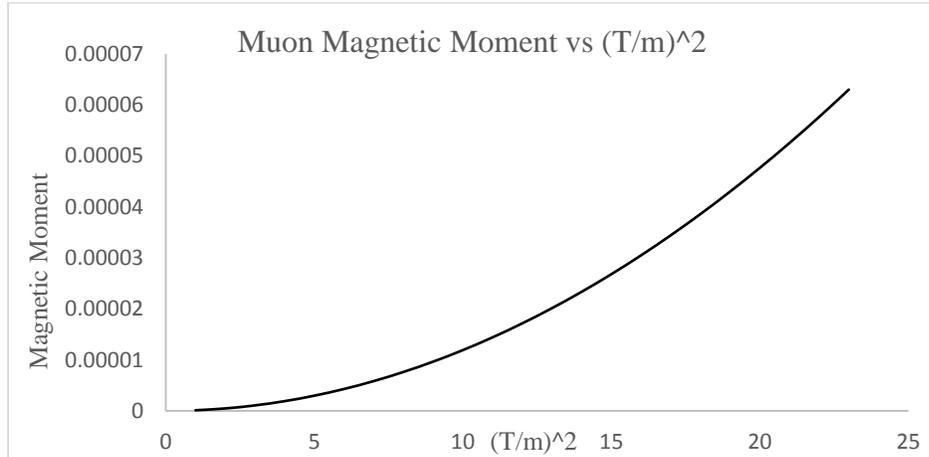

**Figure 3c: Tau Magnetic Moment vs $\left(\frac{T}{m}\right)^2$**

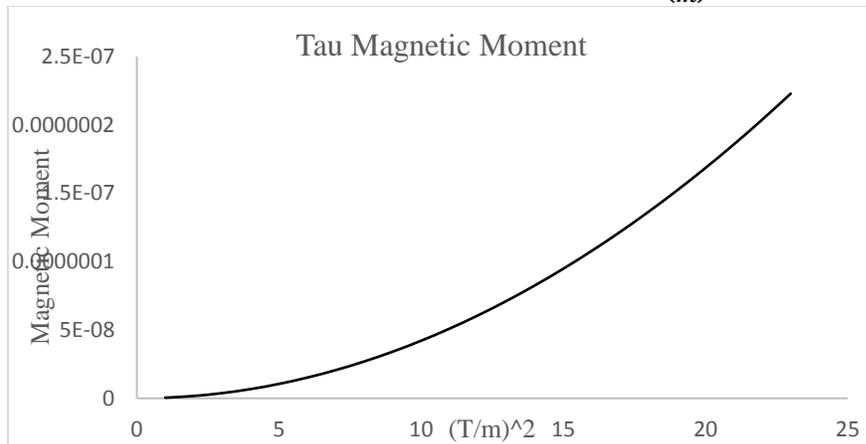

Fig. (3a) shows how the magnetic moment flips when the temperature gets close to 0.22 MeV. This flip will occur for muons and tauons at much higher temperatures. Thermal contributions to the magnetic moment for the muon and tauon particles are too small to make a difference near nucleosynthesis and the concentration of heavier leptons in the universe was ignorable.

In Figs. (3b) and (3c), we see similar magnetic moment behavior of muons and tauons as that of electrons, but their contribution will not be ignorable at larger T. However, electron magnetic moment calculations will need to incorporate weak interactions at larger temperatures.

The graph in Fig. (4) shows a comparison of thermal contributions to the magnetic moment of all three charged leptons.

**Figure 4: Charged Lepton Magnetic Moment vs $\left(\frac{T}{m}\right)^2$**

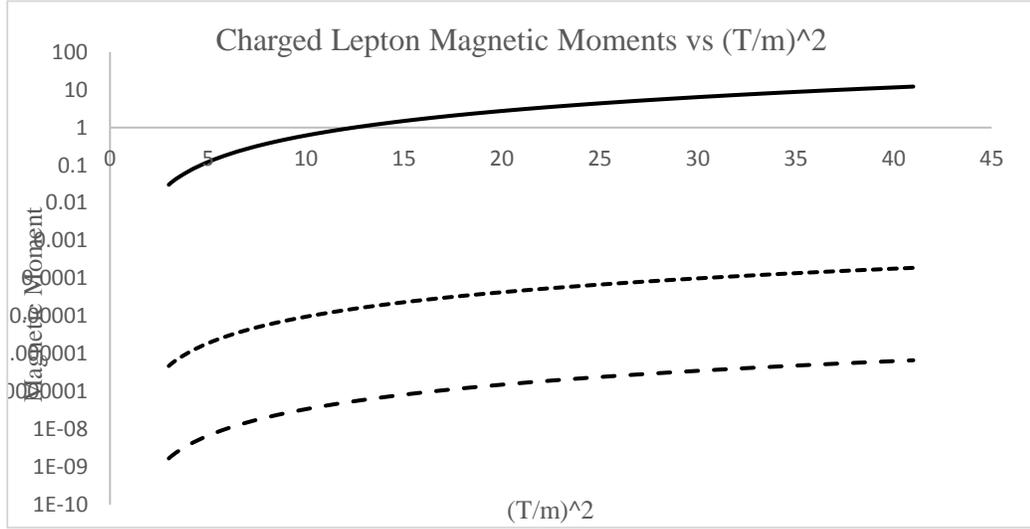

Leading order thermal contribution to leptons near the nucleosynthesis temperature (T ~ $m_e$) is given in Table 1. This is not the exact value as we did not include a, b, c functions contributions. The a, b, c functions can contribute slightly to electron magnetic moment but the order of magnitude comparison is not changed.

**Table 1: Magnetic Moment of Charged Leptons around Nucleosynthesis in the Universe**

| Charged Leptons | Mass (eV) | Corresponding Temperature (K) | Magnetic Moment at T=0 ($\mu_B$) | Thermal Contribution at T=$m_e$ |
|---|---|---|---|---|
| $e$ | $0.511 \times 10^6$ | $0.592 \times 10^{10}$ | 1 | $-7.6 \times 10^{-3} \mu_B$ |
| $\mu$ | $105.65 \times 10^6$ | $0.122 \times 10^{13}$ | $4.8 \times 10^{-3}$ | $-1.19 \times 10^{-7} \mu_B$ |
| $\tau$ | $1776.82 \times 10^6$ | $0.206 \times 10^{14}$ | $2.8 \times 10^{-4}$ | $-4.2 \times 10^{-11} \mu_B$ |

The neutrino mass and the associated magnetic moment are all very small but they are relevant to understand the particle behavior in the early universe. It is known that the magnetic moment of neutrino does not help to resolve the solar neutrino problem. However, its importance cannot be denied in the early universe and inside stars.

We have plotted the magnetic moment of all flavors of neutrinos to compare their contributions in Fig. (5) at the same temperatures. The graphs in Figs. (6) show the relationship between T/m and magnetic moment of individual neutrino flavor. In the beginning it appears to be linear, but its quadratic behavior becomes significant at higher temperatures. These graphs compare the values for each of the three neutrino generations. This makes it easier to see the differences between each particle's magnetic moment with the increase in temperature.

**Figure 5: Neutrino Magnetic Moment vs $\left(\frac{T}{m}\right)^2$**

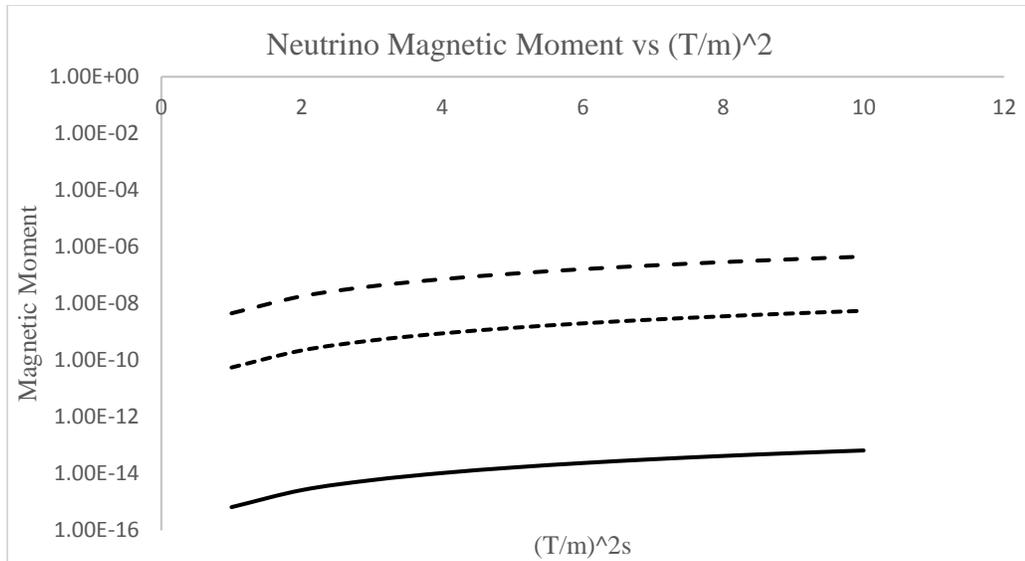

**Figure 6a: Electron Neutrino Magnetic Moment vs $\left(\frac{T}{m}\right)^2$**

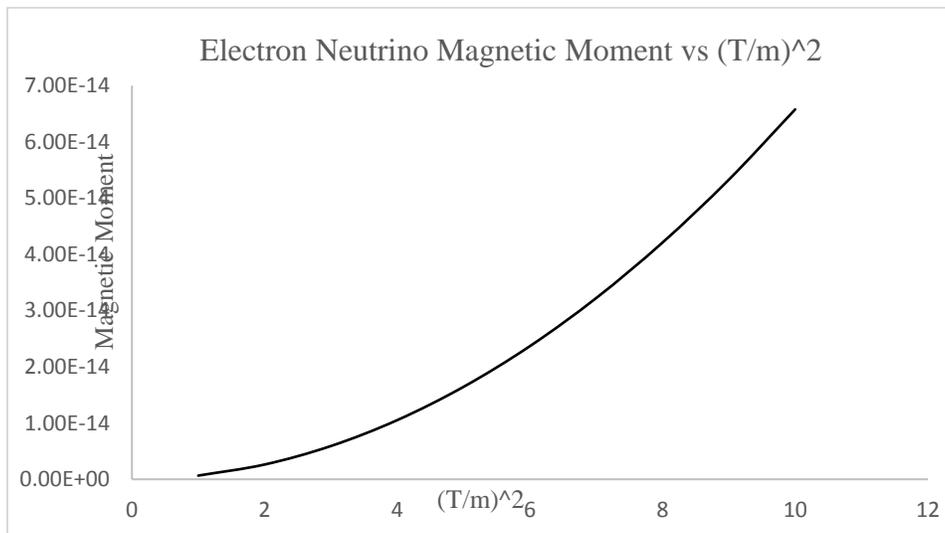

**Figure 6b: Muon Neutrino Magnetic Moment vs $\left(\frac{T}{m}\right)^2$**

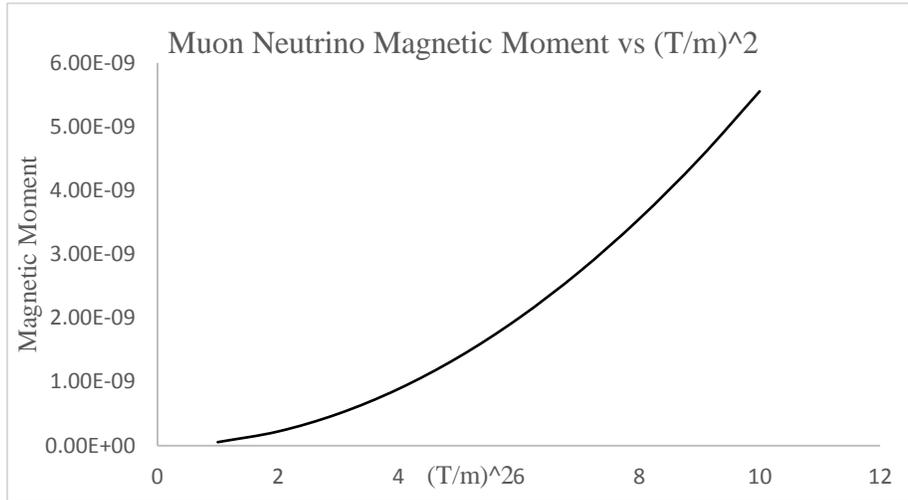

**Figure 6c: Tau Neutrino Magnetic Moment vs $\left(\frac{T}{m}\right)^2$**

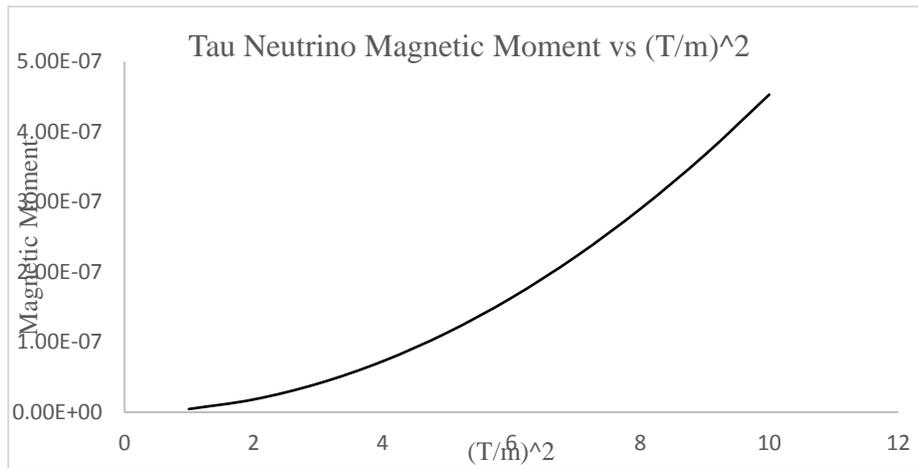

The bubble diagrams in Fig. (1) contribute to magnetic moment at finite temperature. Contribution of tadpole diagrams vanishes in the early universe due to CP symmetry. Table 2 shows that the larger mass of neutrino corresponds to greater magnetic moment. This explains why the tau neutrino has the largest magnetic moment and the electron neutrino has the smallest. This table also shows how the temperature contributes to the magnetic moment.

## Table 2: Magnetic Moment of Neutrinos

| Neutrino Flavor | Mass (eV) | Corresponding Temperature (K) | Magnetic Moment at T=0 $\mu_B$ | Magnetic Moment with Thermal Contribution at T=$m_e$ $\mu_B$ |
|---|---|---|---|---|
| $\nu_e$ | 2.25 | $2.6 \times 10^4$ | $7.2 \times 10^{-19}$ | $6.58 \times 10^{-16}$ |
| $\nu_\mu$ | $1.9 \times 10^5$ | $2.2 \times 10^9$ | $6.08 \times 10^{-14}$ | $5.55 \times 10^{-11}$ |
| $\nu_\tau$ | $1.55 \times 10^7$ | $2.1 \times 10^{11}$ | $4.96 \times 10^{-12}$ | $4.53 \times 10^{-9}$ |

Table 2 gives a quantitative comparison of the magnetic moment near the nucleosynthesis temperatures corresponding to the given values of neutrino masses. These results can easily be modified if the mass limits are further improved in latest experiments.